# Possible evidence for the existence of the Fehrenbacher-Rice band: O $K$-edge XANES study on $Pr_{1-x}Ca_xBa_2Cu_3O_7$

I. P. Hong[1], J.-Y. Lin[2,*], J. M. Chen[3], S. Chatterjee[1], S. J. Liu[4], Y. S. Gou[4], and H. D. Yang[1]

[1]*Department of Physics, National Sun Yat-Sen University, Kaohsiung 804, Taiwan ROC*

[2]*Institute of Physics, National Chiao Tung University, Hsinchu 300, Taiwan ROC*

[3]*Synchrotron Radiation Research Center (SRRC), Hsinchu 300, Taiwan ROC*

[4]*Department of Electrophysics, National Chiao Tung university, Hsinchu 300, Taiwan ROC*

X-ray absorption near edge structure (XANES), resistivity and thermoelectric power have been measured on $Pr_{1-x}Ca_xBa_2Cu_3O_7$. These data reveal an intriguing electronic structure in Pr-doped cuprates. The absorption peak in XANES associated with the Fehrenbacher-Rice (FR) band has been identified. The Ca-doped holes in $Pr_{1-x}Ca_xBa_2Cu_3O_7$ go to both the Zhang-Rice (ZR) and FR bands. Comparative studies on the related samples suggest that the FR band is partially filled and highly localized. Implications of these results on other recent experiments, such as the observation of superconductivity in $PrBa_2Cu_3O_7$ single crystals, are discussed.

Since the discovery of high-temperature superconductors, $PrBa_2Cu_3O_7$ (Pr1237) has stimulated much research interest owing to its unique properties among $RBa_2Cu_3O_7$ ($R$ = rare earths) [1]. In particular, Pr1237 is an insulator and not superconducting, unlike other $R$1237 with $T_C \approx 90$ K. Many theoretical models have been proposed to explain the absence of superconductivity in Pr1237 [1,2]. Among them, the hybrid states (the FR band) with $CuO_2$ plane $pd$ and Pr $f$ states proposed by Fehrenbacher and Rice are considered to be most promising to explain many experimental results [3]. This model was further elaborated by Liechtenstein and Mazin to explain the doping effects of $Y_{1-x}Pr_x Ba_2Cu_3O_7$ [4]. On the other hand, superconductivity was reported in $Pr_{1-x}Ca_xBa_2Cu_3O_7$ thin films or polycrystalline samples synthesized under high pressure, with $x>0.3$ [5,6]. In addition, a trace of superconductivity was also reported in undoped Pr1237 thin films [7]. Very recently, bulk superconductivity of Pr1237 single crystals grown by traveling-solvent floating-zone (TSFZ) method has been reported [8]. Intriguingly, $T_C$ of the TSFZ samples can be enhanced from 85 to 105 K under pressure [8]. Because the behavior under pressure is different from that of other $R$123, the possibility of superconductivity due to the FR band in Pr123 has been proposed [9]. Since the proposed FR band is key to many puzzling properties Pr1237, it is desirable to investigate the existence of FR band. Furthermore, it is not clear whether FR band is conducting or localized if it does exist, and this issue is important to the insulating behavior of Pr1237. Though there have been efforts of x-ray absorption and angular resolved photoemission spectroscopy (ARPES) to look for the FR band; however, only vague evidence was reported [2,10,11]. In this Letter, we report the results of the transport property measurements and x-ray absorption near edge structure (XANES) on $Pr_{1-x}Ca_xBa_2Cu_3O_7$. The existence of the FR band was identified and intriguing information about the FR band was obtained from our data.

Polycrystalline $Pr_{1-x}Ca_xBa_2Cu_3O_7$ samples were prepared by the standard solid-state reaction method [12]. X-ray diffraction patterns show a nearly single phase for Pr1237, and Ca-doped samples appeared to have a minor impurity phase of $BaCuO_2$. However, the Rietvelt analysis indicates that the $BaCuO_2$ impurity phase is less than 4% in molar percentage for $x$ up to $x=0.3$. The magnitude and variation of lattice constants a, b, and c with Ca doping (listed in Table I) are in agreement with the previous studies and considered as evidence for a Pr valance close to +3[12,13]. The electrical resistivity ... was measured by the four probe method. The thermoelectric power $S$ was measured by a standard dc method as described in Ref. [14]. The O $K$-edge x-ray absorption spectra were carried out using linear polarized synchrotron radiation from 6-m high-energy spherical grating monochromator beamline located at SRRC in Taiwan. The energy resolution of the monochromator was about 0.1 eV for the O $K$-edge energy range. Details of XANES experiments can be found in Ref. [15]. The saturation (or "self-absorption") effects were corrected for all measured spectra. The spectra were normalized to the tabulated standard absorption cross section in the energy range 600 to 620 eV as in Refs. [2,16].

The measured results of ... and $S$ are shown in Figs. 1 and 2, respectively. Although ... decreases by more than five orders of magnitude at low temperatures with Ca doping, $Pr_{0.7}Ca_{0.3}Ba_2Cu_3O_7$ is still insulating and not superconducting down to 4.2 K. Ca doping also leads to a significant decrease in $S$ and a change of its temperature dependence. These phenomena are well known as an indication of the increase in the hole number in the $CuO_2$ planes [17]. Interestingly, both ...($T$) and $S$($T$) of $Pr_{0.7}Ca_{0.3}Ba_2Cu_3O_7$ are very similar to those of $Y_{0.4}Pr_{0.6}Ba_2Cu_3O_7$, which is just on the blink of superconductivity [18]. Moreover, the O $K$-edge XANES on $Pr_{1-x}Ca_xBa_2Cu_3O_7$ was studied and shown in Fig. 3. For comparison, the spectra of $YBa_2Cu_3O_7$ (Y1237) and $Y_{0.4}Pr_{0.6}Ba_2Cu_3O_7$ are included. It is known that the XANES is a powerful tool to investigate the unoccupied (hole) states in complex materials. In particular, it is capable of giving information about the carrier numbers on specific sites. In principle, the orientation-resolved spectra from single crystals or oriented thin films in principle could provide more information of the electronic structure of anisotropic compounds. However, according to the past abundant works in the literatures [e.g., 2,15,16,20,21] as long as a conclusion is reached with caution by the spectra from polycrystals, it is usually consistent with that by spectra from single crystals or films.

For Y1237 related compounds, the peak centered around 528.4 eV is related to the Zhang-Rice (ZR) band [19], the peak around 529.1 eV is associated with the upper Hubbard band (UHB) and the feature around 527.9 eV is attributed to the electronic states of the $CuO_3$ ribbons [2,15]. Fig. 3 clearly shows that Ca doping leads to a significant increase in the spectral weight of the peak at 528.4 eV, which manifests an increase in the hole number. Intuitively, this increase in the hole number is consistent with the changes of the transport properties shown in Figs. 1 and 2. Normally, an increase in XANES contribution from ZR band would be accompanied by a decrease in the spectral weight of UHB due to the strong correlation effects in the $CuO_2$ planes [2,15]. Although this correlation between the ZR band and UHB can be qualitatively seen in Fig. 3 for $Pr_{1-x}Ca_xBa_2Cu_3O_7$, the reduction of UHB spectral weight is much smaller than expected. A comparison with the spectrum of Y1237 clearly demonstrates this point. Both with comparable peaks at 528.4 eV, UHB of Y1237 is much weaker than that of $Pr_{0.7}Ca_{0.3}Ba_2Cu_3O_7$. On the other hand, $Y_{0.4}Pr_{0.6}Ba_2Cu_3O_7$ has similar transport properties to those of $Pr_{0.7}Ca_{0.3}Ba_2Cu_3O_7$, but with a smaller peak at 528.4 eV and a comparable UHB. These results can not be reconciled with a simple model involving only the ZR band. Therefore, it is tempting to attributed a part of the spectral weight around 528.4 eV in $Pr_{1-x}Ca_xBa_2Cu_3O_7$ to the FR band, which is formed by the hybridization of O $2p$ and Pr $4f$ orbits and presumably has no correlation with UHB. In this scenario, some of the Ca-doped holes in $Pr_{1-x}Ca_xBa_2Cu_3O_7$ reside on the FR band, and the other holes go to the ZR band which lead to an expense of UHB. Previous reports have suggested that the energy of the FR band could be only slightly higher (within 0.2 eV) than the ZR band [2,10]. This is consistent with the results in Fig. 3 within the limited energy resolution of the experiments.

The quantitative spectral weights of all three denoted features in Fig. 3 have been estimated and the values were plotted in Fig. 4. These values are consistent with those obtained from corresponding single crystals in the literatures [2,16]. (also see Table II) The doping of Ca in $Pr_{1-x}Ca_xBa_2Cu_3O_7$ does not lead to a significant change of the carrier number in the $CuO_3$ chains, similar to the results in Ca-doped Y1237 [16]. However, it induces an increase of carriers in the ZR and FR bands, together with a decrease in the spectral weight of UHB. It is noted that the $Pr_{0.7}Ca_{0.3}Ba_2Cu_3O_7$ and $Y_{0.4}Pr_{0.6}Ba_2Cu_3O_7$ have the same spectral weight of UHB, while the former has a larger spectral weight associated with ZR+FR band than that of the latter. Since UHB is closely related to the ZR band, it is plausible that $Pr_{0.7}Ca_{0.3}Ba_2Cu_3O_7$ has the same number of carriers in the ZR band as that of $Y_{0.4}Pr_{0.6}Ba_2Cu_3O_7$, and the rest of carriers go into the FR band. This scenario is supported by the fact that both samples have almost identical transport properties.

To further discuss the electronic structure, we fix the number of holes in Y1237 to one as in Refs. [2,16]. Using this scaling factor, the hole distribution and the spectrum weight of UHB for several compounds are listed in Table II. Since a 530 eV peak is assigned to a transition into the O(4)-Cu(1)-O(4) dumbbell of oxygen deficient samples [15,20,21], the absence of this feature indicates that all the samples in this experiment are very close to full oxygenation. A minor deviation of the oxygen content from seven would not affect any main conclusion in this Letter. The interplay between ZR, FR and UHB is clearly seen from the Table II. A comparison between $Pr_{0.7}Ca_{0.3}Ba_2Cu_3O_7$ and $Y_{0.4}Pr_{0.6}Ba_2Cu_3O_7$ has been addressed as above. The comparison between $Pr_{0.7}Ca_{0.3}Ba_2Cu_3O_7$ and $Y_{0.6}Pr_{0.4}Ba_2Cu_3O_7$ is another intriguing example. The former has more total carriers than the latter;

however the former is not superconducting and the latter has a $T_C$~45 K. This contrast can be explained by fewer ZR holes in $Pr_{0.7}Ca_{0.3}Ba_2Cu_3O_7$ than in $Y_{0.6}Pr_{0.4}Ba_2Cu_3O_7$. It is also noted that the UHB spectral weight of $Pr_{0.7}Ca_{0.3}Ba_2Cu_3O_7$ is larger than that of $Y_{0.6}Pr_{0.4}Ba_2Cu_3O_7$. By the same reason, it can be understood why $Pr_{0.7}Ca_{0.3}Ba_2Cu_3O_7$, though with a comparable amount of carriers as that of Y1237, is still not superconducting.

According to the calculations in Ref. [4], the FR band grabs 0.22 holes in $Y_{0.4}Pr_{0.6}Ba_2Cu_3O_7$. This would leaves 0.34-0.22=0.12 holes in the ZR band (see Table II). Assuming there are no ZR holes in Pr1237, it gives as many as Ca-doped 0.43-0.26–0.12=0.05 holes going to the FR band in $Pr_{0.7}Ca_{0.3}Ba_2Cu_3O_7$. However, considering the possibility that there might be a few ZR holes left in Pr1237, this hole number is probably underestimated. Therefore, it is very likely that a significant amount of the Ca-doped 0.17 holes go to the FR band in $Pr_{0.7}Ca_{0.3}Ba_2Cu_3O_7$. This conclusion is consistent with a partially filled FR band suggested by ARPES experiments [10]. Furthermore, the observed increase in Pr valence in $Pr_{1-x}Ca_xBa_2Cu_3O_7$ can be explained by doping holes into the FR band [13]. That Ca-doped holes in $Pr_{1-x}Ca_xBa_2Cu_3O_7$ go to both the ZR and FR bands also suggests that the high energy states of the ZR band overlap the low energy states of the FR band. Considering that the $Pr_{0.7}Ca_{0.3}Ba_2Cu_3O_7$ and Y1237 have almost the same number of holes in the ZR+FR bands but with the totally contrast transport properties, we conclude that the FR band is highly localized with little contribution to transport properties.

If all the Ca-doped holes went to the ZR band, $Pr_{0.7}Ca_{0.3}Ba_2Cu_3O_7$ probably would have been a superconductor. However, the share of the additional carriers with the FR band makes 30% Ca doping not enough to bring Pr1237 to superconductivity until 50% Ca is doped [5,6]. Our results also have implications on the recently reported superconductivity in Pr1237 single crystals [8]. There has been proposed possibility that superconductivity found in TSFZ Pr1237 is due to Ba doping on Pr sites [22]. In fact, we have conducted experiments on $Pr_{1-x}Ba_{2+x}Cu_3O_7$. Although the results is not so conclusive as those for $Pr_{1-x}Ca_xBa_2Cu_3O_7$ due to the higher impurity level, the physical properties of $Pr_{1-x}Ba_{2+x}Cu_3O_7$ are similar to those of $Pr_{1-x}Ca_xBa_2Cu_3O_7$, except with an enlarged lattice [23]. Thus, superconductivity due to Ba doping on the Pr sites is in principle possible. On the other hand, since our results suggest a highly localized FR band, the unusual pressure effects in superconducting Pr1237 are difficult to be explained by the FR band contribution. Our results can not explain the reported superconductivity in $Pr_{1+x}Ba_{2-x}Cu_3O_7$, either [24]. Anyway, another study on the same compounds did not find superconductivity [25].

It is noted that the theoretical understanding of the rate of spectral weight is rather limited. Therefore, it is up to what the experimental data reveal. The rate of spectral weight transfer of $Y_{1-x}Ca_xBa_2Cu_3O_y$, which is similar to $Pr_{1-x}Ca_xBa_2Cu_3O_y$ studied in this paper, has been *experimentally* observed in [16]. The effects of Ca doping on the rate of spectral transfer was found to be as expected as by the charge transfer model or the electronic structure calculations. Based on these facts, it is believed the present discussions about the correlation between the ZR band and UHB is plausible. To be cautious, the conclusion is reached by further comparing the spectral results with the transport property measurements. In addition, a comparison to the well studied system $Y_{1-x}Pr_xBa_2Cu_3O_y$ was made. These comparative studies on several sets of data effectively support our point of view.

In conclusion, combination of XANES and the transport properties proves to reveal fruitful understanding of the proposed FR band. The absorption peak in XANES associated with the FR band has been identified with the position close to that with the ZR band. The FR band is highly localized and seems to be partially filled. These results enable us to understand most of the previous experimental findings of superconductivity in Pr related compounds, while leave others unexplained.

We would like to thank C. T. Chen for indispensable discussions. Technical help from P. Nachimuthu, C. W. Chen, and H. H. Li is appreciated. This work was supported by National Science Council of Republic of China under contract Nos. NSC89-2112-M-110-043 and NSC89-2112-M-009-052.

**Table Captions**

Table I.    Lattice parameters of $Pr_{1-x}Ca_xBa_2Cu_3O_7$.

Table II.    The normalized hole distribution and the spectral weight of UHB for various related compounds.

**Figure Captions**

Fig. 1.    Resistivity $\rho(T)$ of $Pr_{1-x}Ca_xBa_2Cu_3O_7$ ($0 \leq x \leq 0.3$). Inset: the same resistivity shown by the logarithmic scale.

Fig. 2.    The thermoelectric power $S(T)$ of $Pr_{1-x}Ca_xBa_2Cu_3O_7$.

Fig. 3.    O $K$-edge XANES for $Pr_{1-x}Ca_xBa_2Cu_3O_7$ and related compounds. Note the relation between the ZR+FR and UHB peaks in different samples.

Fig. 4.    Ca doping $x$ dependence of the spectral weight of the three denoted peaks for $Pr_{1-x}Ca_xBa_2Cu_3O_7$. For comparison, the results of $Y_{0.4}Pr_{0.6}Ba_2Cu_3O_7$ were also plotted at the position of $x=0.3$.

| $x$ | $a$ (nm) | $b$ (nm) | $c$ (nm) |
|---|---|---|---|
| 0 | 0.3870(1) | 0.3924(1) | 1.1703(2) |
| 0.1 | 0.3868(2) | 0.3916(2) | 1.1687(7) |
| 0.2 | 0.3873(2) | 0.3905(2) | 1.1674(7) |
| 0.3 | 0.3874(2) | 0.3895(2) | 1.1668(4) |

Table I.   Lattice parameters of $Pr_{1-x}Ca_xBa_2Cu_3O_7$.

|  | ZR+FR (holes/unit cell) | $CuO_3$ ribbon (holes/unit cell) | UHB (Mbarn/unit cell) |
|---|---|---|---|
| $Pr_{0.7}Ca_{0.3}Ba_2Cu_3O_7$ | 0.43 | 0.55 | 2.27 |
| $Y_{0.4}Pr_{0.6}Ba_2Cu_3O_7$ | 0.34 | 0.52 | 2.25 |
| $Y_{0.6}Pr_{0.4}Ba_2Cu_3O_7$ | 0.39 | 0.50 | 2.06 |
| $YBa_2Cu_3O_7$ | 0.45 | 0.55 | 1.67 |
| $PrBa_2Cu_3O_7$ | 0.26 | 0.53 | 3.03 |

Table II.   The normalized hole distribution and the spectral weight of UHB for various related compounds.

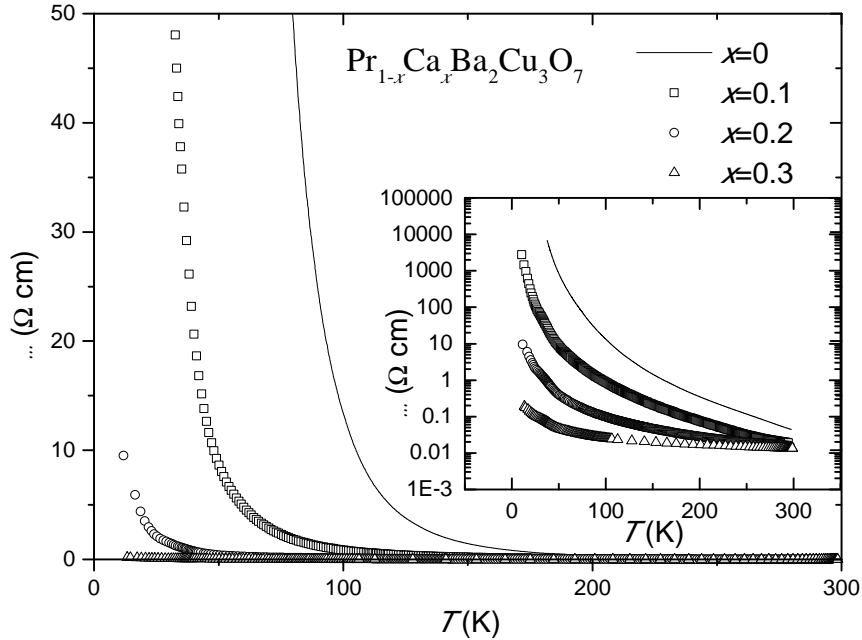

Fig. 1    I. P. Hong et al.

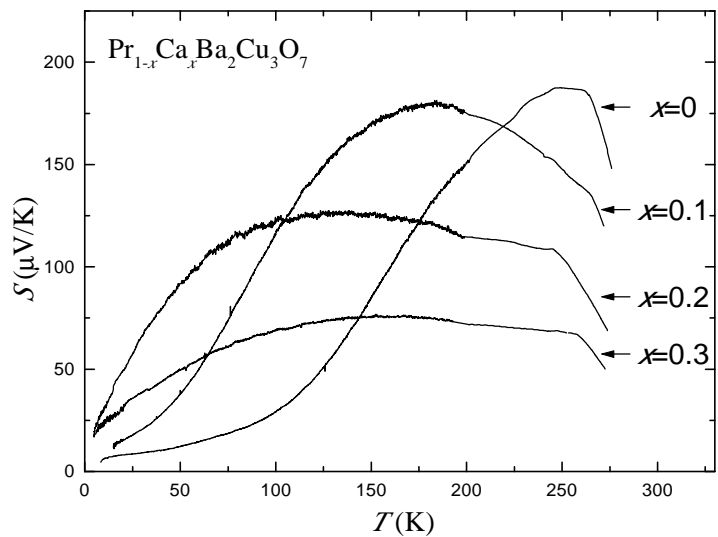

Fig. 2    I. P. Hong et al.

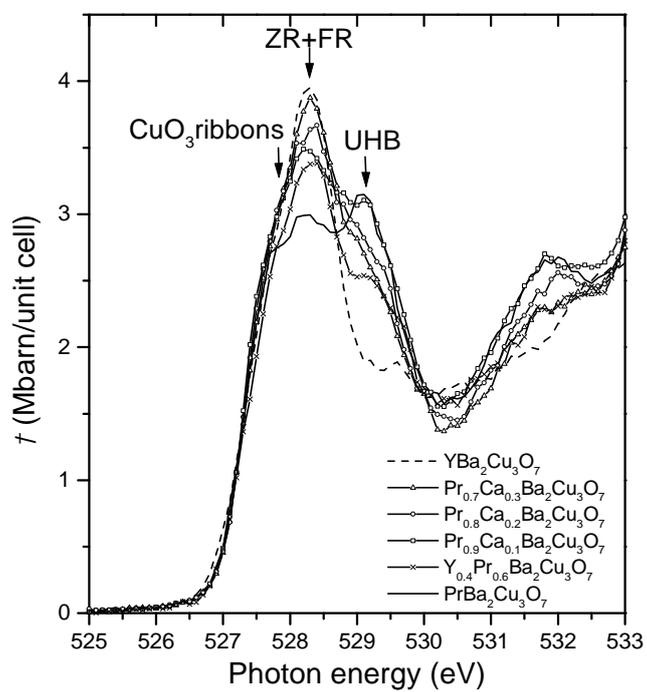

Fig. 3    I. P. Hong et al.

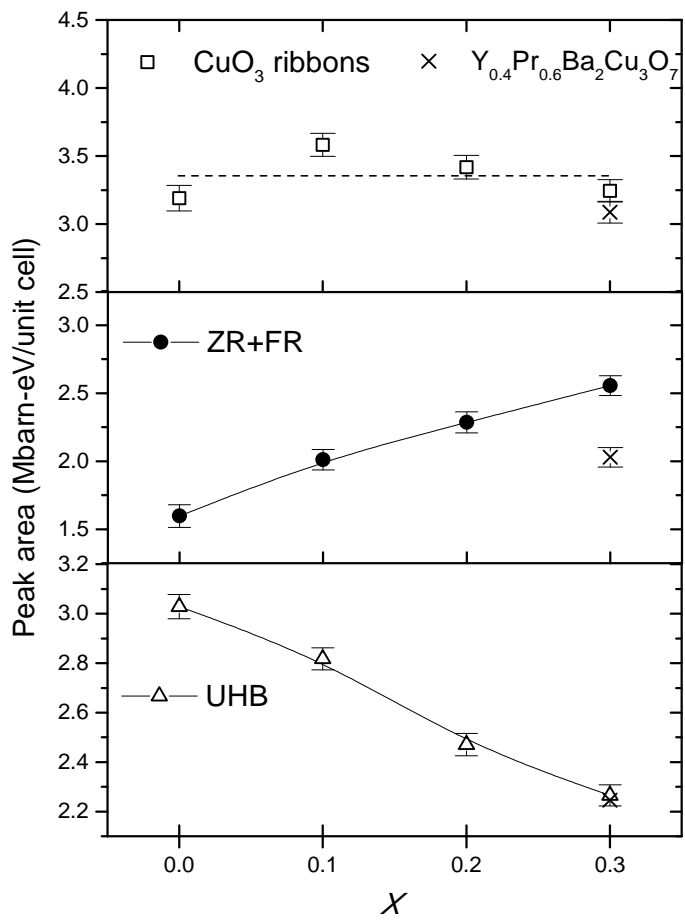

Fig. 4  I. P. Hong et al.